# Scale-up of Dry Impregnation Processes for Porous Spherical Catalyst Particles in a Rotating Drum: Experiments and Simulations

*Yangyang Shen[1], William G. Borghard[1], Bryant Avila[2], Hernan A. Makse[2], and Maria Silvina Tomassone[1] (*)*

*[1] Department of Chemical and Biochemical Engineering*
*Rutgers, the State University of New Jersey*
*98 Brett Rd. Piscataway, NJ 08854, United States*
*[2] Levich Institute and Physics Department*
*City College of New York*
*New York, NY 10031, United States*

(*) Corresponding author
Tel.: +1 848 445 2972
E-mail address: silvina@soe.rutgers.edu

**Abstract**

Catalyst impregnation is one of the most crucial steps for preparing industrial catalysts. The inter-particle variability of the impregnated liquid inside the particles significantly affects the activity and selectivity of the catalyst. Current scale-up practices lead to poor fluid distribution and inhomogeneity in the liquid content. This work aims to understand the dynamic behavior of the particles under the spray nozzle and to develop a scale-up model for dry impregnation processes. We considered five dimensionless numbers in the scaling analysis. We performed Discrete Element Method simulations and experiments of dry impregnation inside porous particles for different vessel sizes. The water content of the particles was compared for other times locations, and the relative standard deviation (RSD) was calculated from the axial water content. Simulation and experimental results show excellent agreement, confirming that the scale-up rules apply to all vessel sizes.

## 1. Introduction

In the manufacturing of heterogeneous catalysts, impregnation is the process of a solution containing active metal components being added to the porous catalyst supports [1-3]. It is desirable to obtain a uniform fluid distribution and similar metal loading in each catalyst particle/pellet after the impregnation step. Impregnation is commonly conducted in a rotating vessel, while the metal solution is sprayed on the support particle [4]. Different processing equipment includes drums, V-blenders, and double-cone blenders. After knowing the impregnation process at the laboratory and pilot scales, it is crucial to reproduce the same quality catalysts at the industrial scale. However, procedures for catalyst manufacturing are usually developed empirically, and current scale-up practices lead to poor fluid distribution and inhomogeneity in metal content.

Scaling up a batch process has long been a puzzle to all chemical engineers [5]. More than half a century ago, a principle of similarity was first proposed for the scale-up of any chemical or physical process [6]. One major prerequisite of this approach is that the physics of the process must be clearly understood [7, 8]. The application of dimensional analysis [9] is also standard practice in the chemical industry and has been successfully implemented for the scale-up of power requirements for mixing equipment [10]. The method [11] involves producing dimensionless numbers and deriving functional relationships that characterize the process entirely. Buckingham's pi theorem is then used to identify the number of dimensionless groups that describe the process. Going from one side of the equipment to the next, the operating parameters are controlled so that the dimensionless numbers are kept constant. In addition, geometrical and dynamic similarities are required to apply the dimensional analysis [11].

Numerous studies have been done on the scaling up of mixing, and coating processes, such as powder blending [12, 13], wet granulation [14-17], pharmaceutical coating [18-20], and spray drying [21], but none of these studies were performed in porous particles. In the area of powder mixing scale-up, Wang and Fan applied the principle of similarity and investigated the scale-up of tumbling mixers [12]. For the scale-up of the dry-particle blending process, Muzzio et al. offered some simple guidelines for free-flowing and cohesive materials [13]. These studies do not account for the spray-related processes. On the other hand, previous works on the scale-up of wet granulation mostly applied a dimensionless spray flux number [14, 15, 22]. Over the past decade, several researchers have investigated various factors that affect the pharmaceutical coating of drug tablets [23-34]. None of these studies mentioned above deals with porous particles.

The discrete element method (DEM) has been applied to investigate particle mixing and segregation [35] and to study the scale-up rules in mixers [36]. In our previous work [37, 38], an algorithm for the spray and inter-particle transfer of fluid onto and within a rotating bed of granular catalyst support was implemented in the DEM simulations. The simulations were validated by experiments utilizing a geometrically identical double-cone blender fixed with a single nozzle impregnator. In this work, we use the same algorithm for water transfer.

It is crucial to be able to reproduce the same quality catalysts at the industrial scale after having knowledge about the impregnation process at the laboratory and pilot scale.

However, procedures for catalyst manufacturing are usually developed empirically, and current scale-up practices lead to poor fluid distribution and inhomogeneity in the metal content. In this work, we present a dimensional analysis and the dimensionless numbers used to characterize the scale-up of the impregnation system. Section 3 contains the description of the simulation method. In Section 4, we present the experimental method, which includes experiments in a small and large cylindrical vessel with validated simulations. These experimental results prove that the dimensionless numbers are suitable scale-up parameters. In Section 5, we present further proof that the dimensionless numbers are proper scale-up parameters using simulations on a portion of a cylinder: a cylindrical "slice" of three different sizes. Finally, the conclusions from the scale-up studies are presented in Section 6.

## 2. Dimensional analysis

We performed a dimensional analysis of the impregnation process. Geometric and dynamic similarities are necessary to apply dimensional analysis. Two systems are considered similar if they are geometrically and dynamically similar. Geometrically similarity is represented by the same shape between different scales and the same ratio of characteristic linear dimensions. Kinematic and dynamic similarity requires that the velocities and forces in all directions have the same ratio. Given the scale-up model, several crucial operation parameters, such as rotation speed and flow rate, can be predetermined to achieve controlled mixing and content uniformity across various scales. Dimensionless numbers are defined, and functional relationships are derived to completely characterize the process.

Extensive studies have been conducted on the scaling of rotating drums due to their simple geometry and wide application in the chemical and process industries as mixers, dryers, kilns, and reactors. Ding et al. developed scaling relationships for rotating drums by non-dimensionalizing the differential equations governing the behavior of the solids motion [39]. Their most relevant dimensionless groups include Froude number, geometric drum ratios, drum fill percentage, and size distribution of the solid particles.

A cylindrical drum is considered our impregnation vessel. A schematic of the vessel with a spraying nozzle is shown in Figure 1. For this system, we considered five different dimensionless numbers for scale-up analysis. All these quantities were derived taking into account that the system of different scales should have a geometric and kinematic and dynamic similarity. The dimensionless numbers were kept constant for different scales.

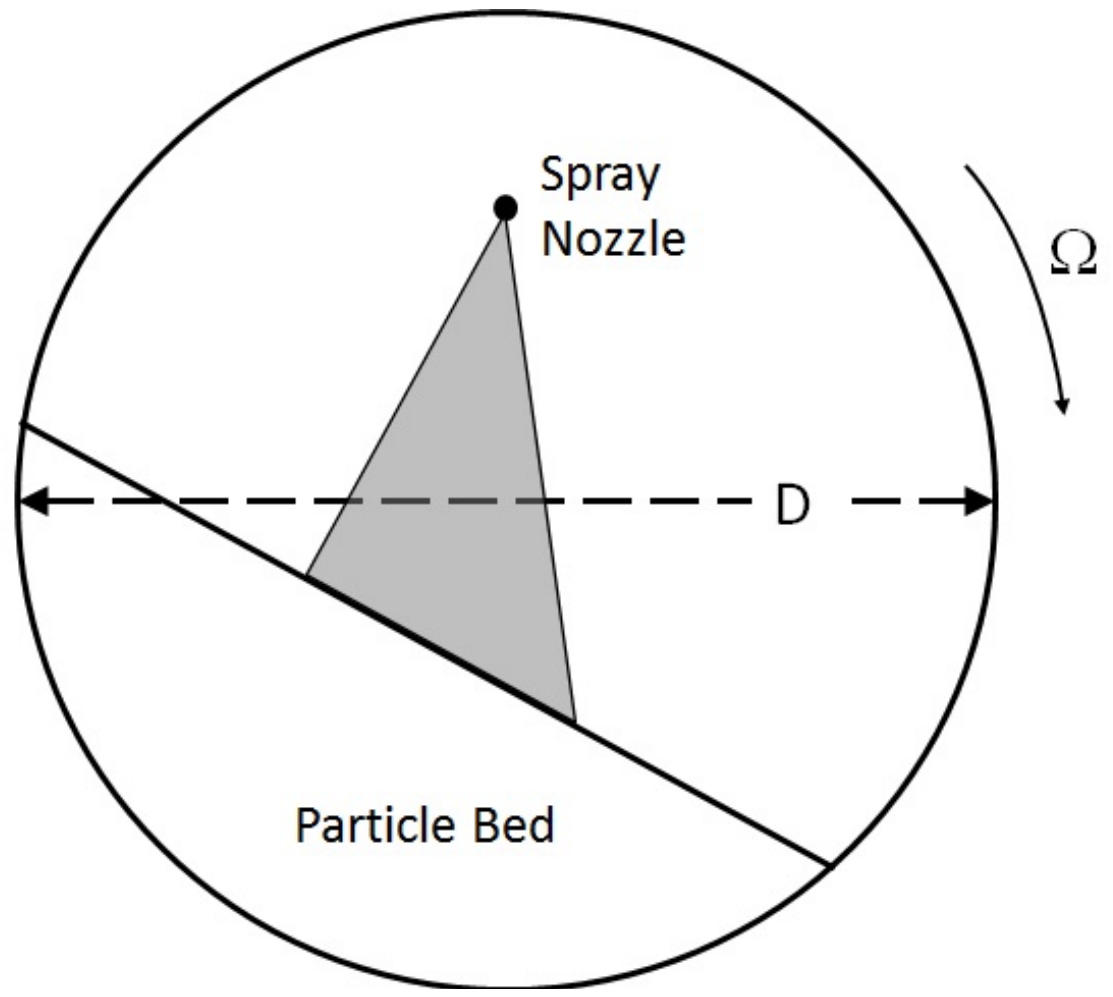

**Figure 1.** A schematic of the impregnation process in a rotating drum

$\Pi_1 = \frac{L}{D}$ , $\Pi_2 = \frac{A_{wetted}}{L^2}$ , $\Pi_3 = Fr = \frac{\Omega^2 \times D}{2g}$ , $\Pi_4 = C_Q = \frac{Q}{\Omega \times D^3}$ and $\Pi_5 = \mathrm{FL} = \frac{4\mathrm{M}^{\mathrm{total}}}{\rho_{\mathrm{p}} \varphi \pi \mathrm{D}^2 \mathrm{L}}$

where L is the length of the cylinder, D is the diameter of the cylinder, $A_{wetted}$ is the wetted area of the cylinder, $Fr$ is the Froude Number, $C_Q$ is the Flow Rate Number, $FL$ is the fill level, $\rho_p$ is the particle density, $\varphi = 0.69$ is the packing fraction of the spherical particles, Ω is the angular velocity (or rotational speed), and $Q$ is the flow rate of spraying solution. Other researchers have designed similar dimensionless numbers. Ban et al. [34] defined a dimensionless number as $\frac{X_{center}}{D}$, which includes the spray center location ($X_{center}$) and the diameter of the cylinder (D). We instead defined $\frac{A_{wetted}}{L^2}$. Additionally, the wetted area in their work is a square, but the wetted area, in our case, is circular. The calculation of each dimensionless quantity converts specific operation parameters into a term value, and by keeping the value fixed in all scales, we are able to derive the operation parameters from one scale to another scale. Notice that to have a geometrically similar impregnator, the fill level is also kept constant, and consequently, the mass of the material also scales up as $M_2 = M_1 \frac{D_2^2 \cdot L_2}{D_1^2 \cdot L_1}$. M is the loading mass, D and L are the diameter and length of the drum, respectively.

The Froude number gives a correlation between vessel size and rotational speed of the vessel. Within different $Fr$ regimes, different types of bed motion have been identified [46], such as slipping, rolling and cascading. The rolling bed is preferred, providing favorable conditions, such as steady state and a flat bed profile for mixing and uniformity. For a given vessel size, if the rotational speed and Froude number are too small, the system does not move or move like a cradle, and no mixing occurs. At huge Froude numbers, the vessel rotates at high speed, causing particle collision and attrition, which are to be avoided during the impregnation process. An optimal regime of the Froude number can be found when the system shows the best mixing behavior. A more detailed analysis of the Froude number is given in section 5.2.

The flow rate number ($C_Q$) compares the rate of spray droplets ($Q$) being deposited onto the particle bed and the rate of particles being refreshed at the bed surface. The latter is correlated with the rotational speed (Ω) and the vessel diameter ($D$). In general, smaller spray rates tend to give more homogeneous liquid distribution during and after the total liquid is sprayed. However, the process takes longer. Using more significant spray rates, the spraying can be completed in shorter times, but the particles show more significant inhomogeneity of the fluid content in the particle bed. An optimal value for the Flow Rate number allows the impregnation process to finish in an acceptable time frame and the final product to have a uniform distribution of the metal solution. However, the process conditions may be limited to the equipment settings.

## 3. Simulation method

Discrete element method (DEM) has been increasingly used to study granular materials and particle systems. In the DEM, the motion of individual particles is computed according to the Newton's second law of motion. The motion of a solid particle is expressed by

$$ma = \sum F_{Contact} + \sum F_{Body} \qquad [1]$$

where m and $a$ are the mass and acceleration of a solid particle, respectively. The term $\sum F_{Contact}$ accounts for all the normal and tangential contact forces, which are due to particle-particle or particle-boundary collisions. $\sum F_{Body}$ denotes the sum of all body forces due to gravity.

DEM simulations in this work were performed using the EDEM commercial software package developed by DEM Solutions, Ltd., based on an original method proposed by Cundall and Strack [40]. The contact forces are calculated using Hertz-Mindlin no-slip contact model. It is based on a soft contact model or elastic approach, in which the magnitude of the repulsive force is related to the amount of overlap. The normal force is calculated using a damped Hertzian normal contact model [41] with the damping term given by Tsuji et al. [42]. The magnitude $F^n$ from a contact that resulted in a normal overlap $\delta_n$ is given by:

$$F^n = -k_n \delta_n^{3/2} - \gamma_n \dot{\delta}_n \delta_n^{1/4} \qquad [2]$$

where $k_n$ is the Hertzian normal stiffness coefficient, $\delta_n$ is the deformation (normal particle overlap), $\gamma_n$ is the normal damping coefficient, and $\dot{\delta}_n$ is the rate of deformation.

In the above equation, $k_n$ is obtained by:

$$k_n = \frac{4}{3} E_{eff} \sqrt{R_{eff}} \qquad [3]$$

$E_{eff}$ is the effective Young's modulus of two colliding entities (two particles or a particle and a wall). For entities with Poisson's ratios $\nu_1$ and $\nu_2$, Young's moduli $E_1$ and $E_2$, $E_{eff}$ is given by:

$$E_{eff} = \frac{1-{\nu_1}^2}{E_1} + \frac{1-{\nu_2}^2}{E_2} \qquad [4]$$

$R_{eff}$ is defined as the effective radius of the contacting particles. In case of a particle–wall collision, the effective radius is simply the particle radius. While in the case of particle–particle collision, with the two contacting particles having radii $R_1$ and $R_2$, the effective radius is obtained by:

$$R_{eff} = \frac{R_1 \times R_2}{R_1 + R_2} \qquad [5]$$

With knowledge of the normal stiffness coefficient and a chosen coefficient of restitution ε, the normal damping coefficient is calculated as:

$$\gamma_n = 2\sqrt{\frac{5}{3}\left[\frac{ln(\varepsilon)\times\sqrt{mk_n}}{\sqrt{ln^2(\varepsilon)+\pi^2}}\right]} \qquad [6]$$

Following the work of Mindlin and Deresiewicz [43], the tangential force $F^t$ is calculated in a similar method as its normal counterpart. The tangential contact force also consists of elastic and damping components. When a tangential overlap of δt is detected, and there is a corresponding normal overlap of $\delta n$ due to the same contact, then the tangential force is calculated as:

$$F^t = -k_t \delta_t - \gamma_t \dot{\delta}_t \delta_n^{1/4} \qquad [7]$$

where $k_t$ the tangential stiffness coefficient and $\gamma_n$ is the tangential damping coefficient.

In the above equation, $k_t$ is calculated by:

$$k_t = 8 G_{eff} \sqrt{R_{eff}} \sqrt{\delta_n} \qquad [8]$$

$G_{eff}$ is the effective shear modulus. For two entities with shear moduli $G_1$ and $G_2$:

$$\frac{1}{G_{eff}} = \frac{2-\nu_1}{G_1} + \frac{2-\nu_2}{G_2} \qquad [9]$$

The tangential displacement (or overlap) $\delta_t$ is calculated by time-integrating the relative velocity of tangential impact, $v_{rel}^t$ between two colliding entities (either interparticle or particle–wall contact):

$$\overrightarrow{\delta_t} = \int \overrightarrow{v_{rel}^t} dt \qquad [10]$$

The capabilities of EDEM include user-defined functions and various features for simulating the impregnation process, which has been developed and validated in our previous work [17]. In a typical simulation, initially, catalyst support particles are filled in a blender, and then liquid droplets are continuously released from a nozzle. An algorithm has been developed to allow droplets to transfer their mass to the catalyst support particles when they are in contact. After contact, the droplets disappear.

The fluid spray is modeled as discrete droplets, which are sprayed from a nozzle located above the rotating bed and are absorbed upon contact with the simulated catalyst particles. When the droplet comes into contact with a particle, the fluid droplet essentially disappears while simultaneously transferring its mass to the simulated catalyst support particle, leading to a net increase in the mass. The mass flow rate of the fluid is defined as: $Q_{spr} = NV = N\frac{m}{\rho}$ where N is the number of fluid droplets, V is the volume, m is the mass and  is the density of each droplet. Analogous to the experimental conditions, the particles in this study absorb fluid up to 60% of their weight. After saturation of the catalyst particle occurs, additional fluid allows the support particle to be considered supersaturated, and as a result, transferred excess fluid to any non-saturated particle that it comes into contact with. Our hypothesis in the model is that the amount of fluid transferred between two particles in every contact when one of them is supersaturated, is defined as $Q_{tr} = \kappa(m_i - m_j)$. In this equation, $\kappa$ is a proportionality constant that reflects the rate of fluid transfer, and $m_i$ and $m_j$ are the respective mass of each of the particles; for water transfer $\kappa$ is as defined as 0.01. So, the amount of fluid transferred per contact is a function of the difference in the excess of fluid of the contacting particles. When the amount of fluid absorbed is more than the fluid contained in the pore volume, the fluid transfer algorithm allows the excess of fluid on a specific catalyst support particle to be transferred to another adjacent particle at a specific rate.

In the simulations, a number of parameters are used to accurately represent the experimental system. The values of these initial parameters are listed in Table 1. These material properties are obtained directly from the alumina catalytic support used in the experiments.

The catalyst support particles have a predetermined threshold to mimic the pore volume of the experimental catalyst. When the amount of liquid absorbed is more than this value, the liquid transfer algorithm also allows the excess liquid on a specific catalyst support particle to be transferred to another adjacent particle, at a user-specified rate. In the simulations, a number of parameters were used to accurately represent the experimental system. The values of the material properties were obtained directly from the alumina catalytic support used in the experiments and are shown in Table 1.

**Table 1.** Initial parameters and material properties used in the simulations.

| **Parameter** | **Value** |
|---|---|
| Density of particle | 1500 kg/m$^2$ |
| Diameter of particle | 6.2 mm |
| Shear modulus | $2\times10^6$ N/m$^2$ |
| Poisson ratio | 0.25 |
| Coefficient of restitution | 0.5 |
| Coefficient of static friction | 0.8 |
| Coefficient of rolling friction | 0.1 |

## 4. Experimental method

Water impregnation experiments were performed in 2 different cylindrical sizes: 1) small cylinder ($D_1$=20cm and $L_1$=30cm) and 2) large cylinder ($D_2$=40cm and $L_2$=60cm). Both systems have two spray nozzles located close to each other, generating a continuous wetted area at the center of the vessel. Figure 2 illustrates the experimental set up for impregnation in the cylindrical vessel of 2 sizes. The rotational speed and spray rate were determined by keeping the Froude number (*Fr)* and Flow rate number ($C_Q$) constant for the two scales. After the completion of spraying, additional rotations were also performed. Table 2 lists the process parameters used for small and large cylinders. The Froude number chosen for both cylinders is $Fr = \frac{\Omega^2 \times D}{2g} = 0.0308$, and the Flow Rate number is $C_Q = \frac{Q}{\Omega \times D^3} = 6.54 \times 10^{-5}$.

It is worth noticing that there is not a unique Froude. In our scale-up studies, it is essential to first validate the simulation parameters using experiments and, secondly, verify that the chosen dimensionless numbers are suitable scale-up parameters. This will be presented in the following two sections after describing the experimental setup.

(A)

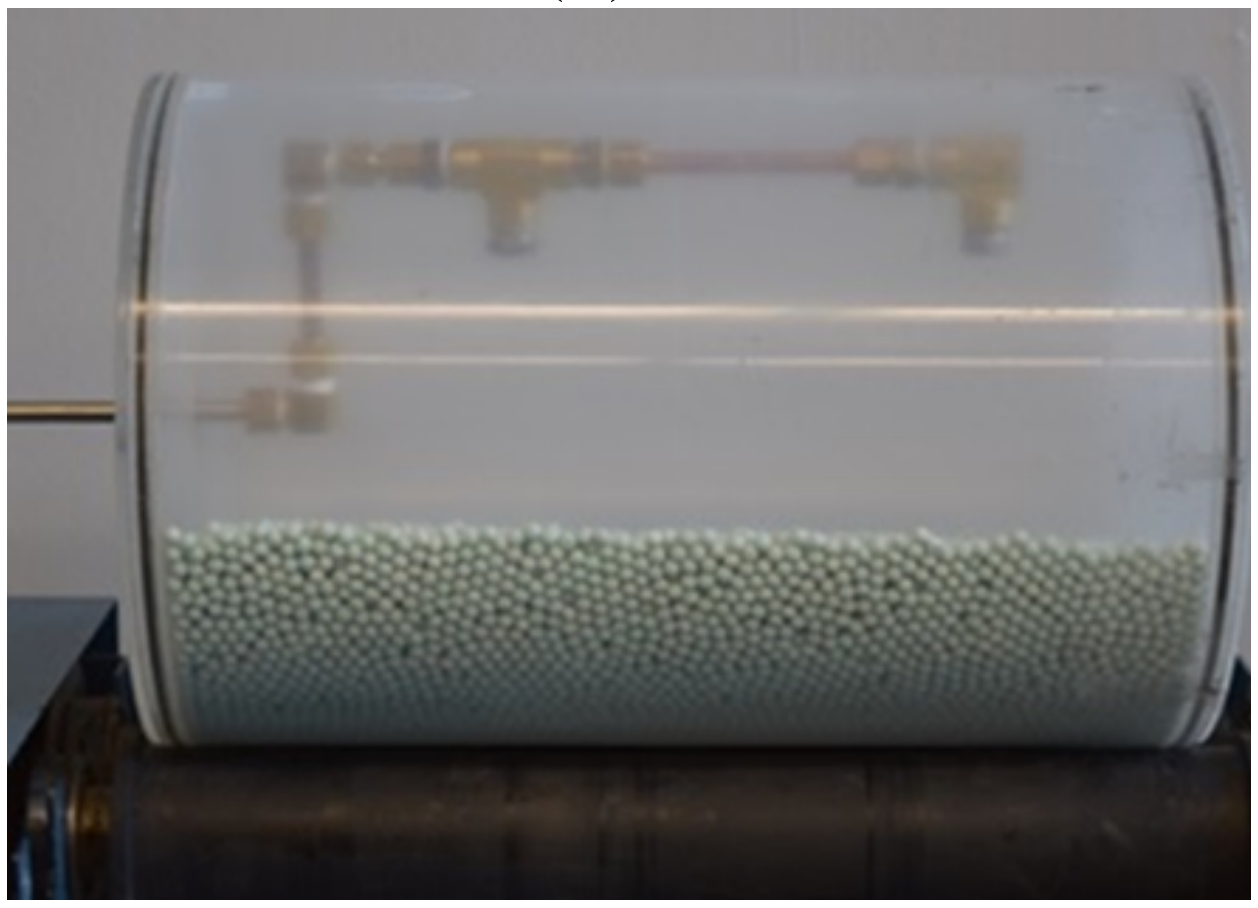

(B)

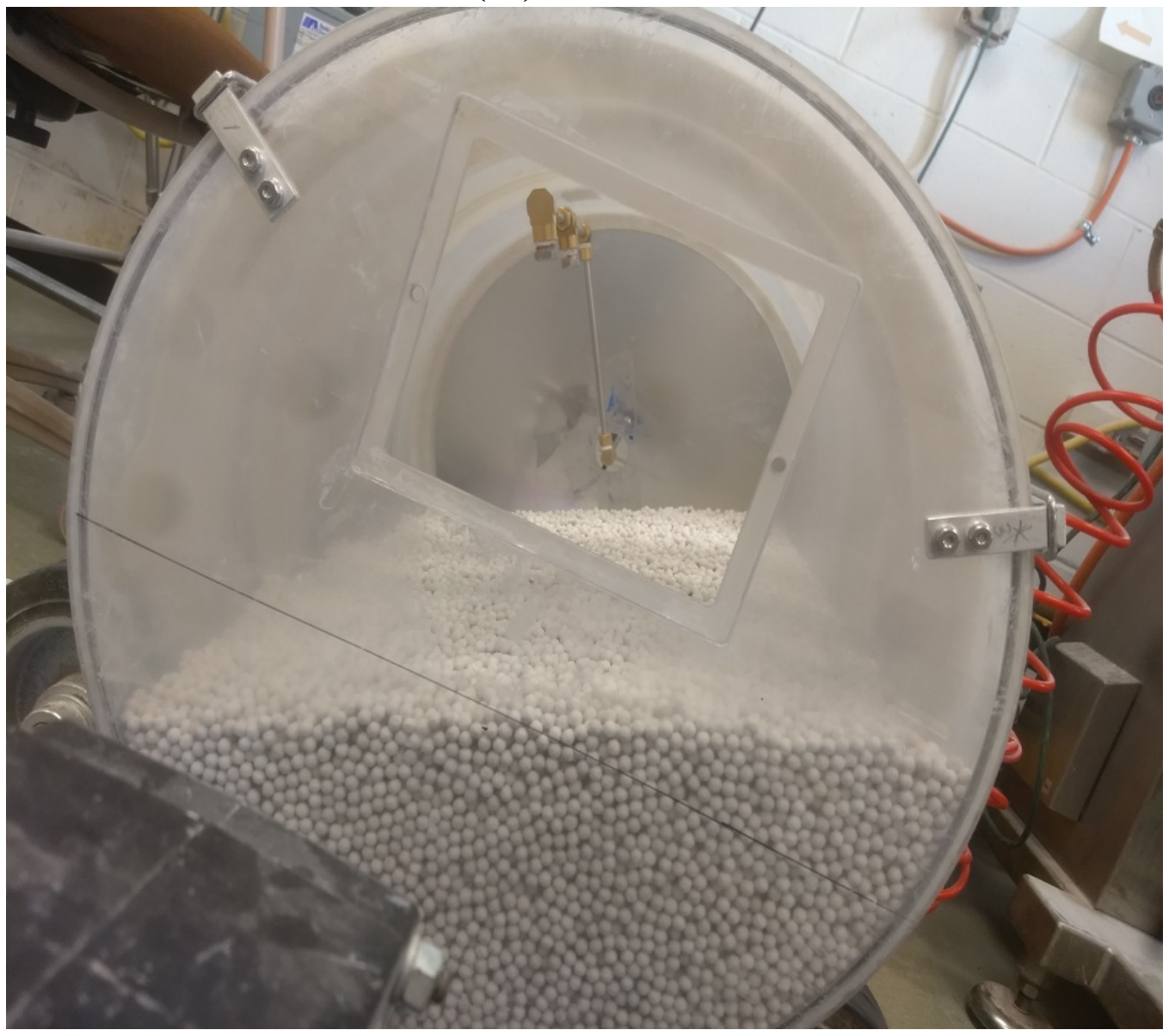

**Figure 2.** Experimental setup for impregnation in a cylindrical vessel with two adjacent nozzles. (A) Small cylinder ($D_1$=20cm and $L_1$=30cm). (B) Large cylinder ($D_2$=40cm and $L_2$=60cm).

**Table 2.** Geometric parameters of the cylinders and process parameters used in the experiments.

| **Small Cylinder** | **Large Cylinder** |
|---|---|
| $L_1$ = 30 cm | $L_2$ = 60 cm |
| $D_1$ = 20 cm | $D_2$ = 40 cm |
| Rotation Speed $\mathbf{\Omega_1}$ = **17 rpm** | Rotation Speed $\mathbf{\Omega_2}$ = **12 rpm** |
| Diameter of one Wetted Area = 7.5 cm | Diameter of one Wetted Area = 15.5 cm |
| Mass $M_1$ = 1.7 kg | Mass $M_2$ = 13.6 kg |
| Volume of water = 1.01L | Volume of water = 8.01 L |
| Flow Rate $\mathbf{Q_1}$ = **3 L/hr** | Flow Rate $\mathbf{Q_2}$ = **17 L/hr** |

Scale-up rules can be used to determine the process parameters for a large vessel ($D_2$, $L_2$) based on those of a small vessel ($D_1$, $L_1$). Based on our chosen dimensionless numbers, we can observe that if the vessel size is increased by a factor of $N$ (i.e. $\frac{D_2}{D_1} = \frac{L_2}{L_1} = N$), then the amount of material increases by cube of $N$ (i.e. $\frac{M_2}{M_1} = N^3$), the wetted area increases by square of $N$ (i.e. $\frac{A_{wetted,2}}{A_{wetted,1}} = N^2$), the rotation speed decreases by square root of $N$ (i.e. $\frac{\Omega_2}{\Omega_1} = \sqrt{N}$), and the flow rate increases by $N$ to the power of 2.5 (i.e. $\frac{Q_2}{Q_1} = N^{2.5}$).

### 4.1 Experimental setup

Experiments were conducted with the vessel loaded at a fixed fill level of 30% by volume corresponding to 1.7 kg and 13.6 kg of alumina spheres, respectively. 6.2mm γ-alumina spheres were purchased from Saint-Gobain Norpro (Stow, OH, USA). The granular spheres contained a surface area of 200 $m^2/g$ and a pore volume of 0.6 $cm^3/g$. An impregnator was retrofitted into the system using Swagelok ¼ in. fittings and a ¼ inch NPT nozzle adapter. The flow was controlled using a Cole-Parmer Masterflex™ peristaltic pump retrofitted to the Swagelok fittings.

During the impregnation process, samples of approximately 20–25 particles were retrieved every 5 min The samples containing 6.2mm impregnated spheres were removed for analysis at 5 points along the axis of rotation at the top of the bed as indicated in Figure 3. All samples were stored in air-tight glass vials prior to analysis for moisture. Moisture was analyzed by heating the samples to 300 °C for 6 h and measuring the associated mass change. Moisture content was normalized by the weight of the sample. The fluid content uniformity was characterized by using the relative standard deviation (RSD), which was calculated from the ratio of the standard deviation (σ) and the average value of the fluid content (C) in the samples of catalytic particles after impregnation. RSD was calculated by the two equations shown below:

$$RSD = \frac{\sigma}{\bar{C}} \qquad [11]$$

and

$$\sigma = \sqrt{\frac{n\sum C_i^2 - (\sum C_i)^2}{n(n-1)}} \qquad [12]$$

In general, lower RSD values indicate less variability between samples, which implies better mixing and fluid content uniformity. There is no consensus regarding a fixed value for the relative standard deviation (RSD) uncertainty to indicate uniformity; rather, the value of the RSD used as a threshold depends on the application and the sample size [44]. In some cases, the observed variability is combined with the observed bias in the sample average to provide a combined criterion [30], so that the limit in RSD depends on the observed level of deviation in the sample mean. The size of our samples is about 20 to 25 particles per sample, and in each experiment, we tested five samples taken at five different times and locations, with three repetitions. Therefore, a total of approximately 1500 units per experiment were tested, which provides enough measurements to yield a normal or near-to-normal distribution. It is standard and widely accepted [44, 45] to consider a 95% or higher confidence interval in normal distributions to assure batch acceptability. Our data show a 95% confidence interval (i.e., two standard deviations from the mean value taken at 100% pore volume filling), and our measurements of the RSD on those samples show RSD values that range from 0.09 to 0.109. Hence, we take RSD=0.1 as our criterion for a reasonably good degree of uniformity since RSD values smaller than 0.1 correspond to a 95% confidence interval. In addition, in general, in the catalysis industry it is widely accepted that a 10% variation in the catalyst quality is a reasonable variability, since other quality factors (surface area, activity, metals dispersion, crush strength, etc.) are often specified to this level of variation.

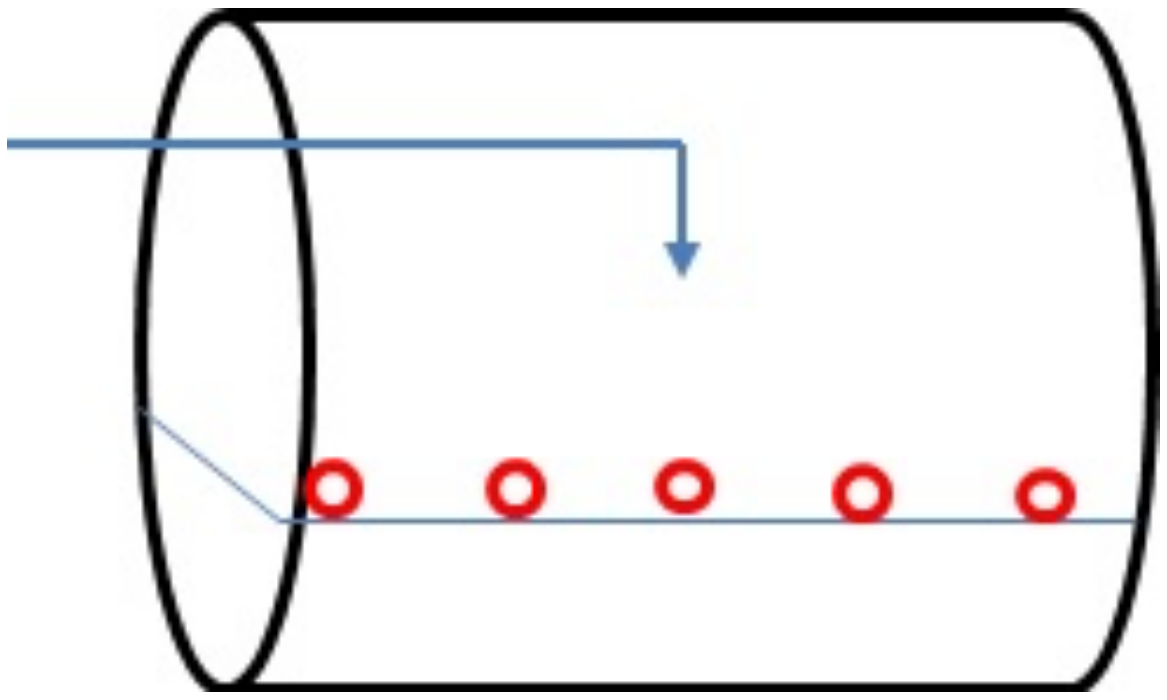

**Figure 3.** Schematic of experimental set up and sampling positions along the rotational axis, shown as the red circles. The arrow indicates the location of the opening from where the samples were removed.

### 4.2. Validation of DEM simulation parameters for both small and large cylindrical vessels

In order to validate the parameters chosen in the DEM model, we performed water impregnation simulations in cylindrical vessels with full length. Two sizes of vessel were considered: a small cylinder ($D_1$=20cm and $L_1$=30cm), and a large cylinder ($D_2$=40cm and $L_2$=60cm). Simulations and experiments have a "one to one" size correspondence (i.e. we kept the exact geometrical dimensions and nozzle configuration as compared to the ones used in experiments). Figure 4 shows the initial setup of the simulation and a schematic of the nozzle position for the small cylinder. The large cylinder has the same setup. Two nozzles are at the axial center of cylinder with the wetted area adjacent to each other. These simulations are computationally very expensive due to the size of the vessel. Also notice that these simulations have axial dispersion as opposed to the cylindrical slice shapes used previously. The particles used in the simulation have a uniform size of 6.2mm.

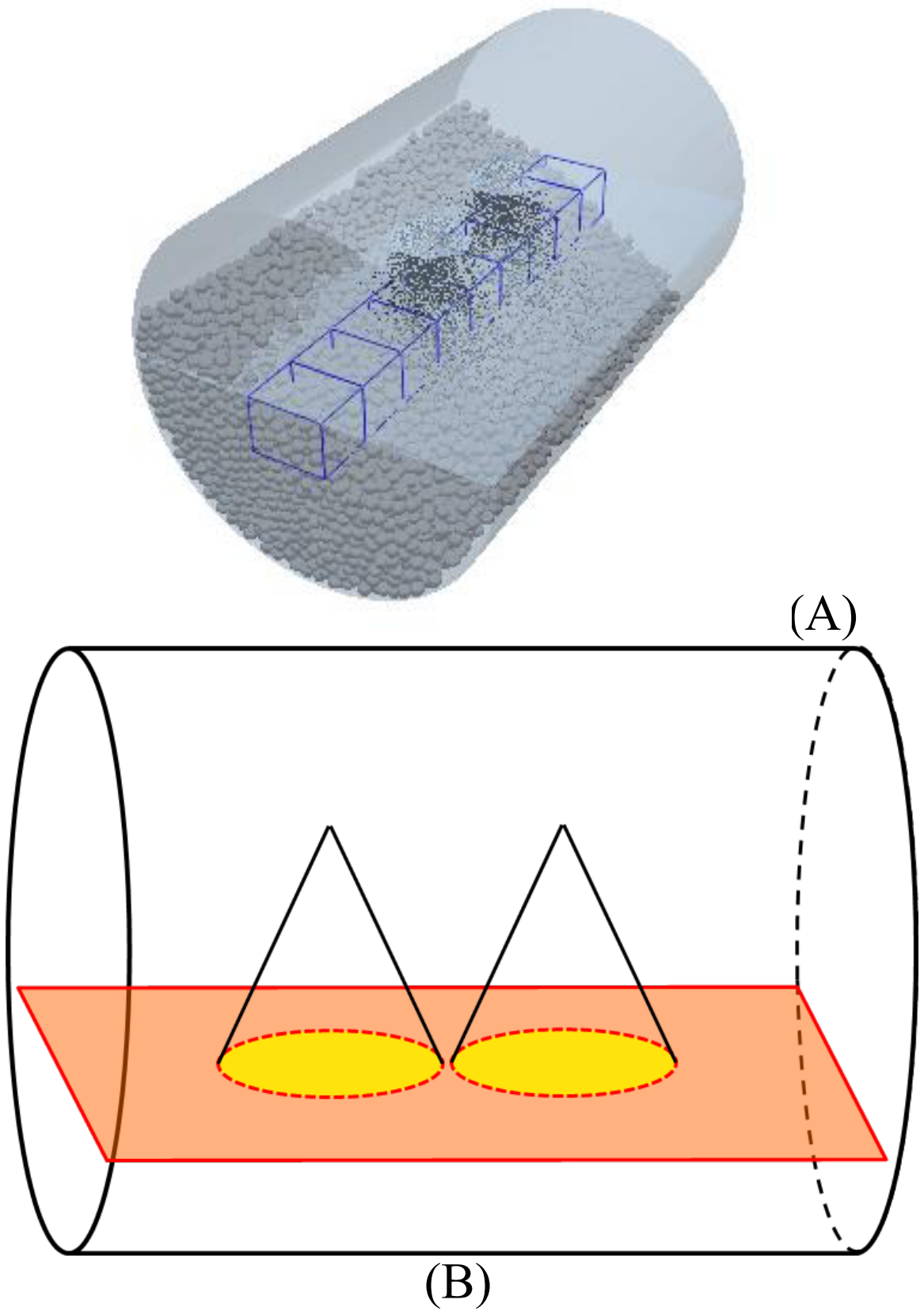

**Figure 4.** (A) DEM simulation setup of the cylindrical vessel (D=20cm, L=30cm). The boxes along the axial direction are the sampling points in the simulation. (B) schematic of the two nozzles in an adjacent position in the cylinder.

The other operation conditions were the same as what were used in the experiments for both size cylinders. The details and geometric parameters are shown in Table 2. The water content in the particles were calculated in the 5 axial positions and compared the results with experimental results. The experiments were conducted in the small and large cylinders that have identical set up. Each experimental point represents the mean value of three experiments; the error bars represent one standard deviation of the mean.

Figure 5 shows the comparison of the water content of the particles along the axial direction at different times for the small cylinder ($D_1$=20cm and $L_1$=30cm). The axial position is marked left to right by LL (far left), L (left), C (center), R (right), and RR (far right); temporal evolution begins at the bottom and increases vertically in 5-minute intervals. It is important to note that the temporal evolution is represented by individual lines vertically, where each line represents 5 minutes of impregnation. During spraying, the axial water content in the particle bed increases at a steady speed, as shown in Figure 5(A). The case of the small cylinder exhibits some anisotropy in the initial 10 minutes of experiments but is improved to a more uniform profile by the completion of the 20-minute impregnation. Good agreement is observed between simulations and experiments, indicating that the parameters chosen in the simulations are validated.
After the spraying, additional rotations are performed in both the simulation and experiment for another 5 minutes. Figure 5(B) shows the water content after additional rotations. The profile at 25 minutes indicates excellent uniformity after additional rotations. The patterns exhibited by the simulations and the experiments are similar, showing that water distribution reached a uniform profile. This indicates that content uniformity can be further improved by additional rotations.

Figure 6 shows the comparison of the water content of the particles along the axial direction at different times for the large cylinder ($D_2$=40cm and $L_2$=60cm). In this case, the water content also shows some disturbance during the entire process of impregnation.

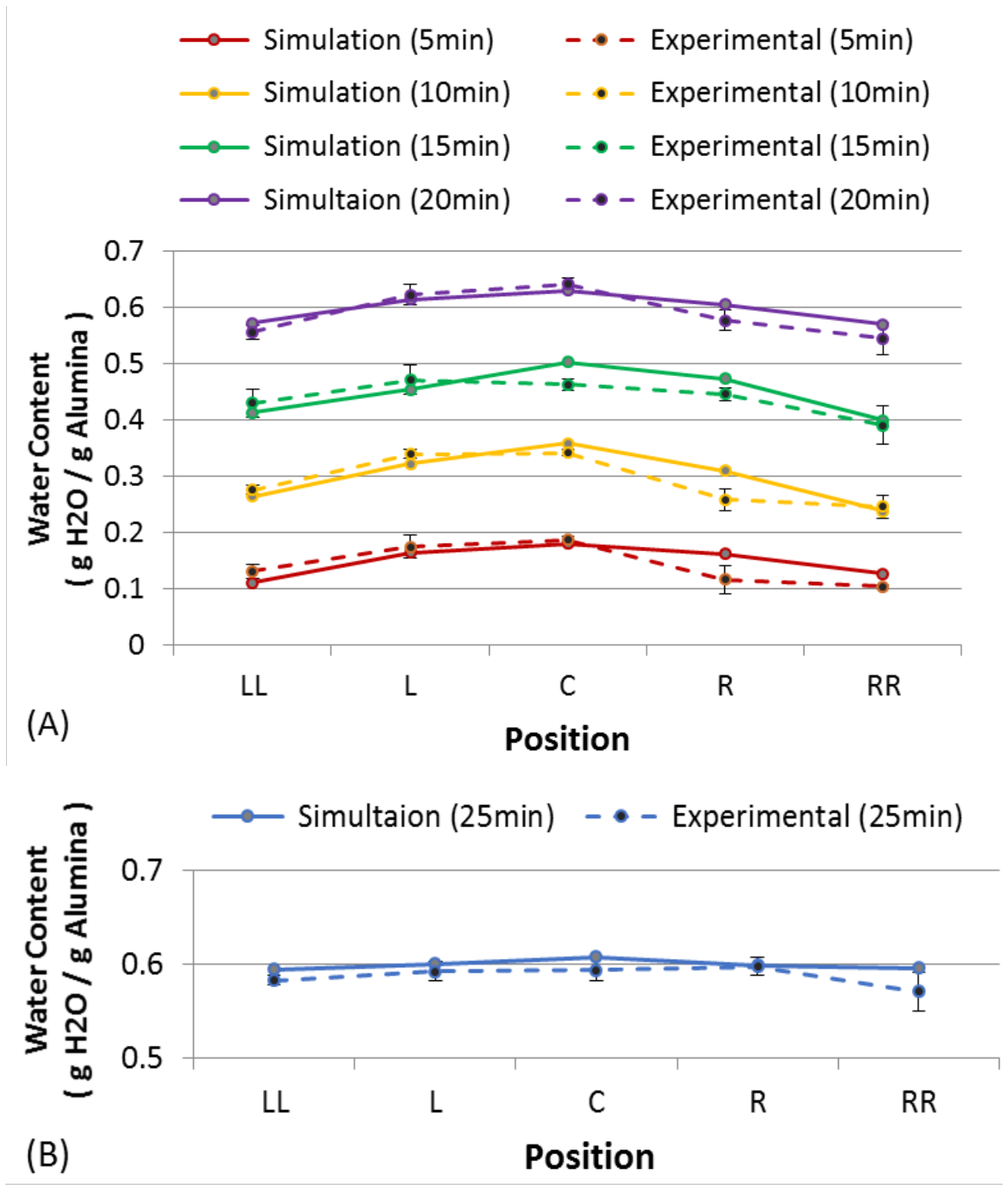

**Figure 5.** Comparison between simulation and experiments of the axial water content distribution for the small cylinder ($D_1$=20cm and $L_1$=30cm) at different times. (A) during impregnation and (B) additional rotations.

This could be due to inconsistency in the sampling positions. Then after additional rotations, the water content exhibits better distribution in the axis at 35 minutes. A non-trivial amount of rotations may be required to reach the same level of uniformity exhibited by the small cylinder case. Good agreement is also observed between simulations and experiments, indicating that the parameters chosen in the simulations are validated for the large cylinder system.

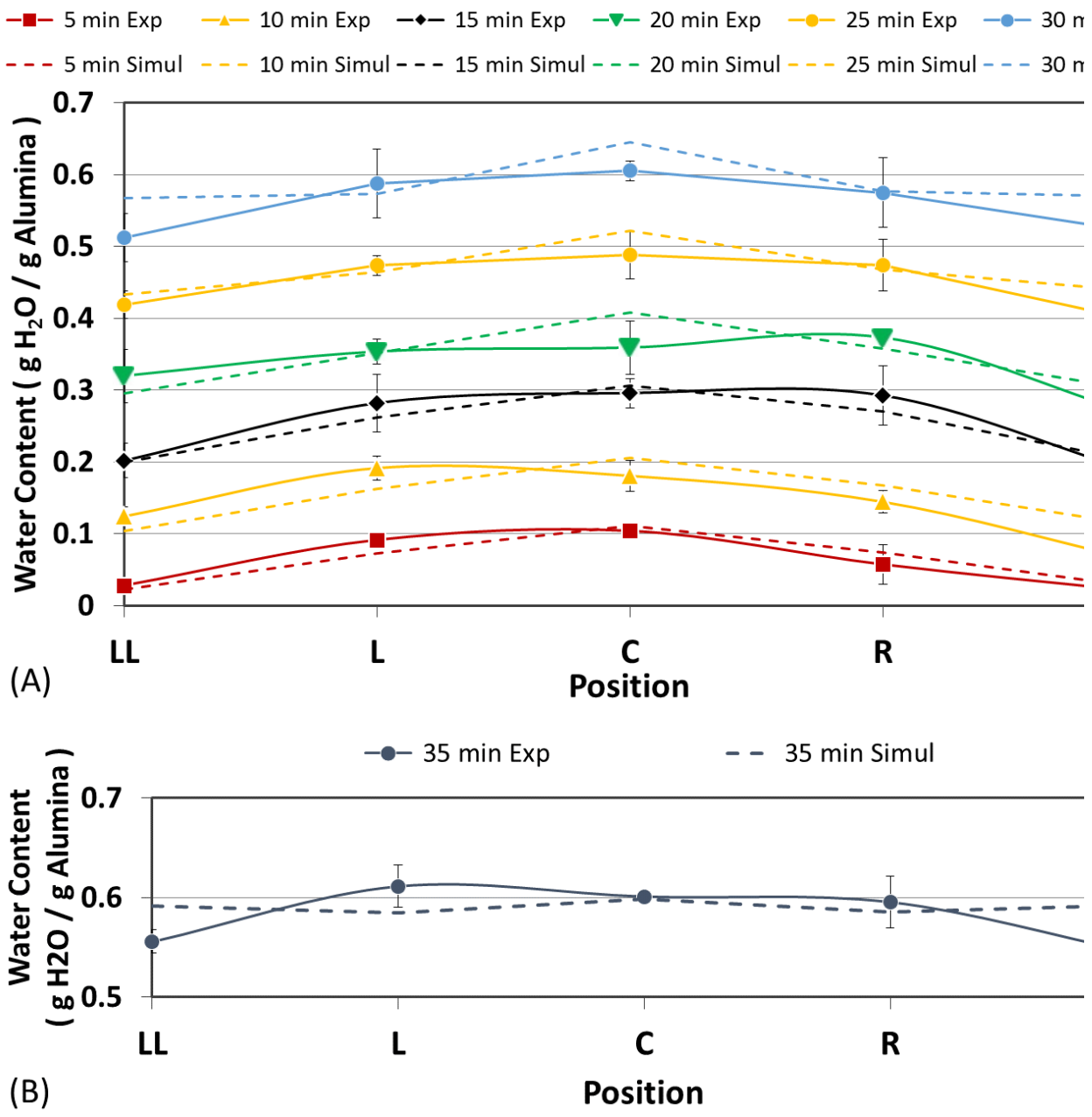

**Figure 6.** Comparison between simulation and experiments of the axial water content distribution for the large cylinder ($D_2$=40cm and $L_2$=60cm) at different times. (A) during impregnation and (B) additional rotations.

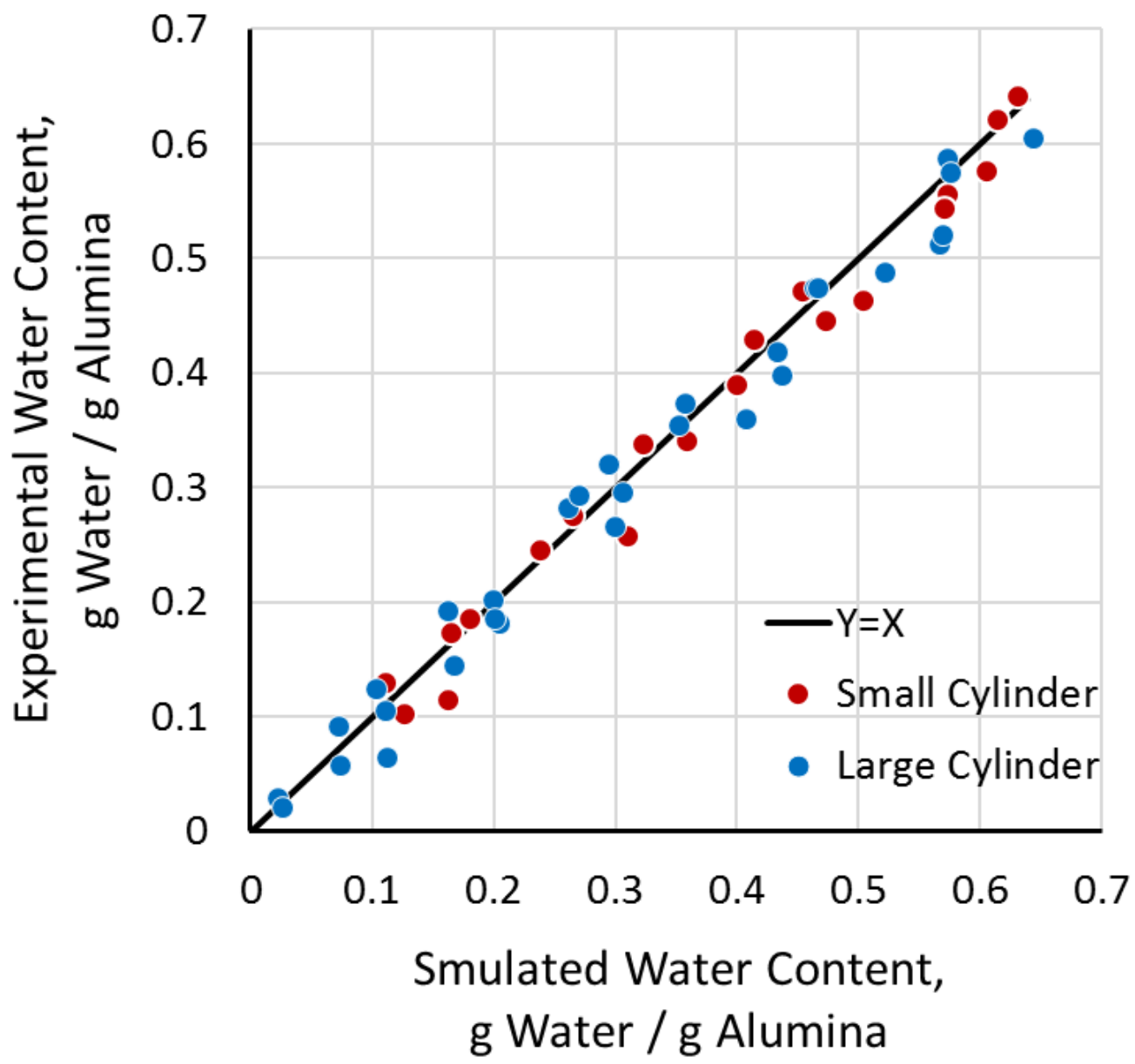

**Figure 7.** Parity plot comparing the experimental values and computed values for the water content in the catalyst support particles. The results from both small and large cylinders are included.

Additionally, this agreement is evident in the parity plot shown in Figure 7, where the experimental and simulation values for water content for both small and large cylinders are shown. These results suggest that the DEM model and chosen parameters were able to successfully describe the dry impregnation process. This allows us to move forward to study the scale-up of the impregnation process using DEM simulations.

The relative standard deviation (RSD) of the water content along the axis were computed and compared between simulation and experiment for both small and large cylinders. Figure 8 shows RSD as a function of time. The results of experiment were averaged between experiments and error bars represent one standard deviation between three experiments. Good agreement between simulation and experiment is also evident in the comparison of RSD.

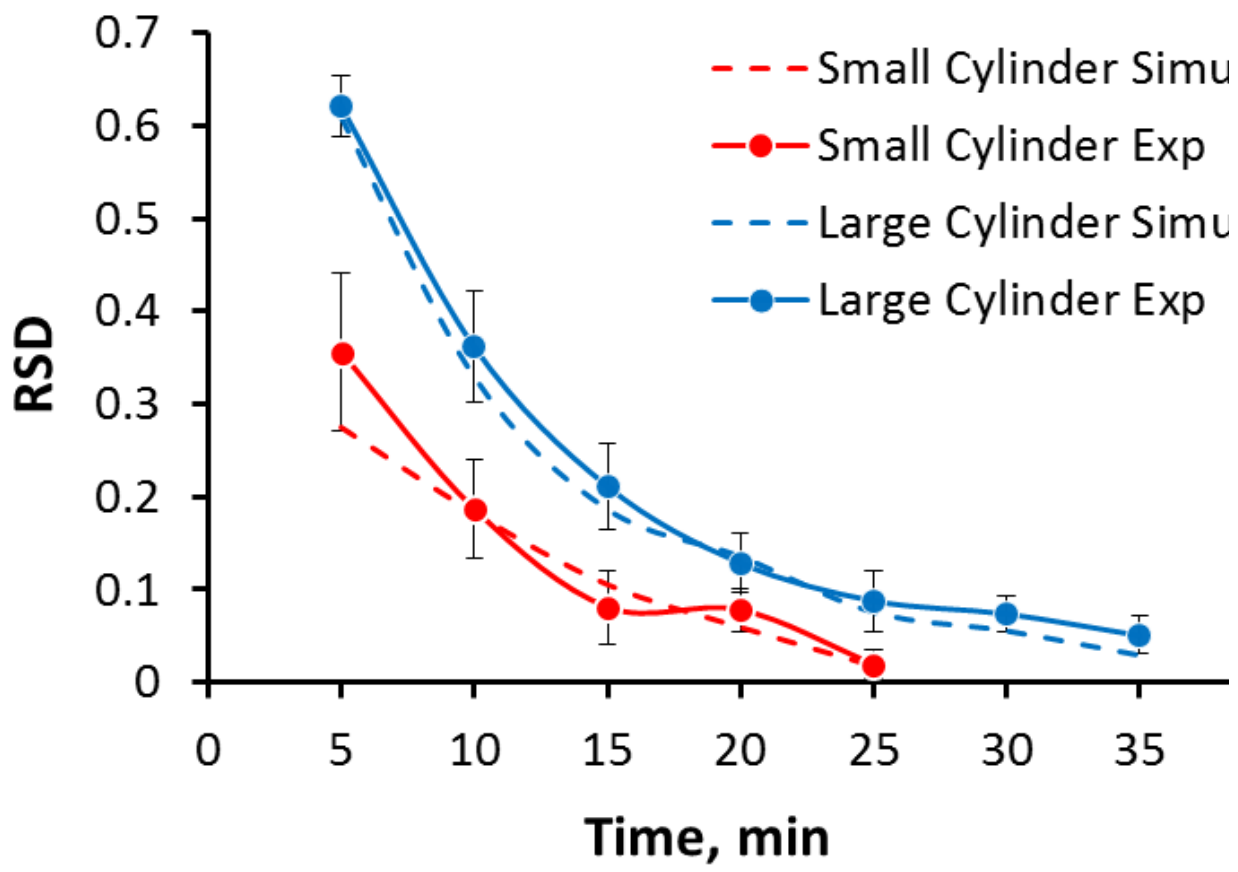

**Figure 8.** The relative standard deviation of the axial water content as a function of time for both small and large cylinders.

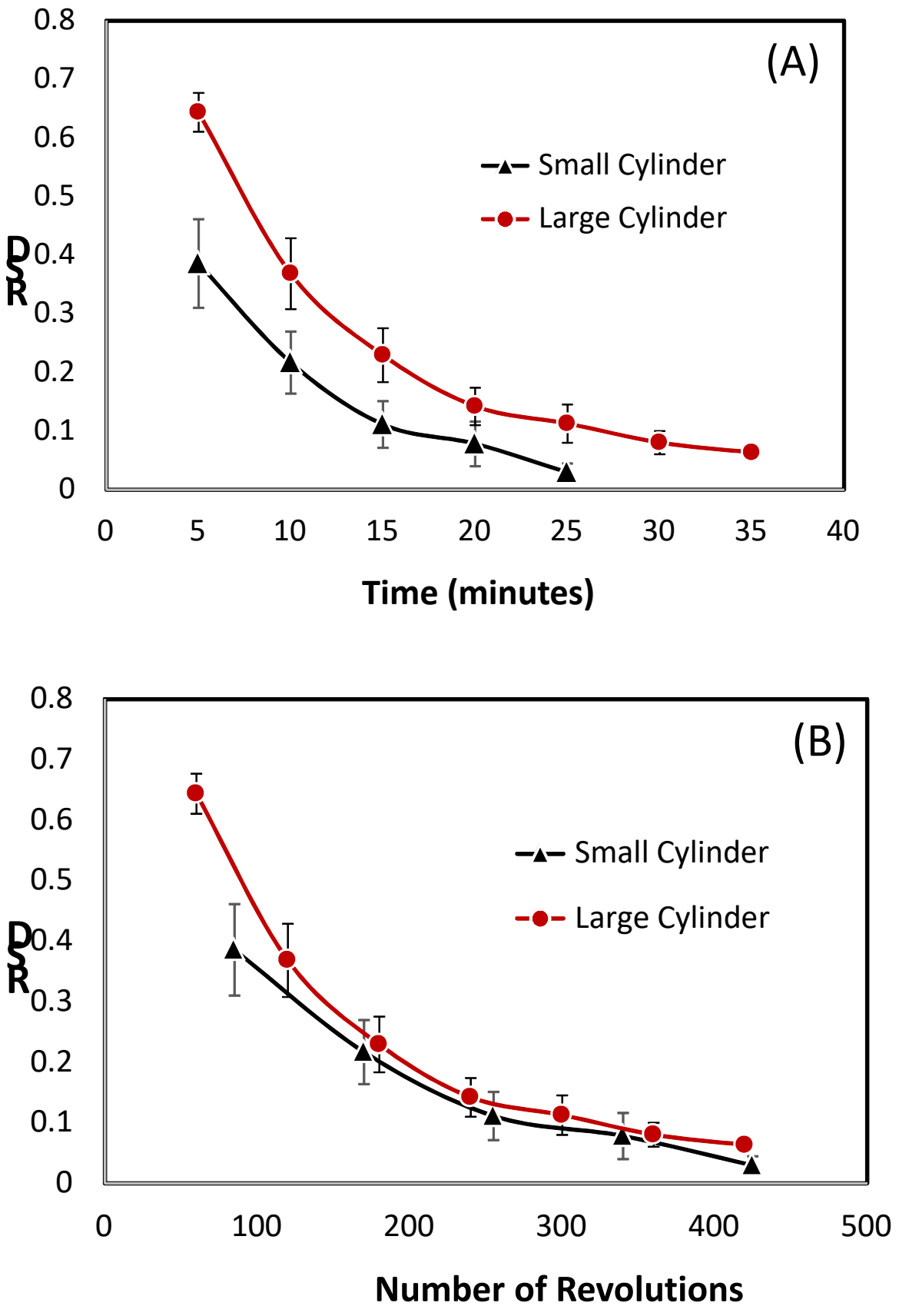

**Figure 9.** Experimental RSD for small and large cylinders: (A) as a function of time and (B) as a function of the number of revolutions.

### 4.3 Verification that dimensionless numbers are suitable scale-up parameters

To verify that the two-size cylinders have the same performance in the impregnation process, we measured the Relative Standard Deviation (RSD). RSD is calculated from the amount of water in the axial direction; it is plotted here as a function of time and the number of revolutions, as shown in Figure 9. Figure 9(A) shows the plot of RSD as a function of time; it is evident that the time needed for RSD to reach 0.1 differs between small and large cylinders. Figure 9(B) shows the plot of RSD as a function of the number of revolutions. It is observed that at the end of impregnation, both cylinders undergo the same number of rotations. In addition, it is shown that the two RSD curves collapse into one between the two different scales within the error bars, indicating that the dimensionless numbers are suitable scale-up parameters.

## 5. Simulation studies on scale-up

**5.1 Details of the Simulation Model:** To further examine these dimensionless quantities, DEM simulations were performed in three sliced cylinders with different diameters with a width covering the entire spraying area. This geometry represents a portion in a cylindrical blender that includes one nozzle in the system. Note that the cylindrical slices with periodic boundary conditions are used to save computational time. The diameter of cylinder was varied from 10cm to 20cm, and 30cm.

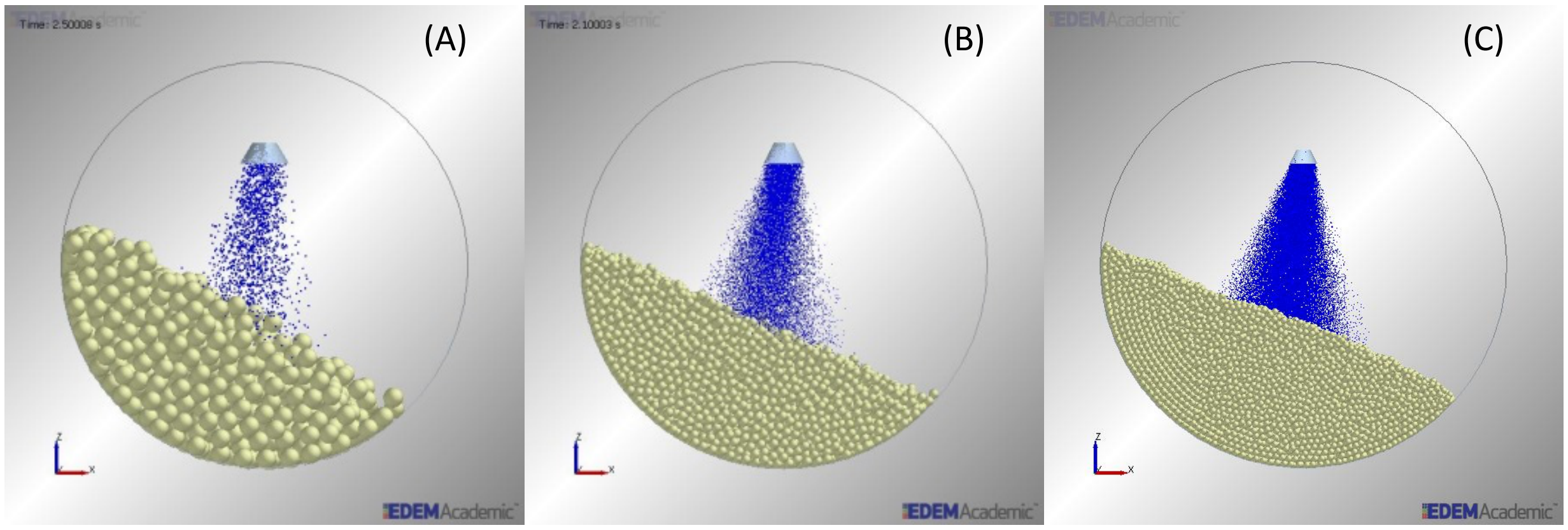

**Figure 10.** Initial simulation setup of the cylindrical slices: (A) D=10cm, (B) D=20cm, (C) D=30cm.

For all three sizes, the total number of particles in the system is determined to have the same fill level in the vessel. The rotational speed and spray rate are calculated from the Froude number (*Fr*) and the Spray Rate number ($C_Q$), respectively. As the vessel size increases, the rotational speed decreases and the spray rate increases. The simulation conditions for different vessel sizes are given in Table 3. The initial setup of simulation is shown in Figure 10. Water content in the particles can be observed from simulation snapshots at different times during the impregnation process, as shown in Figure 11. During the first 1 minute, particles start to receive water from the spray nozzle and show a low water content in particles. Then particles keep adsorbing water and the water content increase with time. A small group of particles could have higher water content then the rest of the particles. At the end of spraying it is observed almost all the particles have similar water content, reaching to 100% pore volume filled.

| | Small slice<br>**$D_1$ = 10cm**<br>**$W_1$ = 4cm** | Medium slice<br>**$D_2$ =20cm**<br>**$W_2$ =8cm** | Large slice<br>**$D_3$ =30cm**<br>**$W_3$ =12cm** |
|---|---|---|---|
| **Number of particles** | $N_1$ = 8000 | $N_2$ = 8000 | $N_3$ = 27000 |
| **Rotational speed (keeping Fr constant)** | $\Omega_1$ = 12.7 rpm | $\Omega_2$ = 9 rpm | $\Omega_3$ = 7.35 rpm |
| **Spray rate (keeping $C_Q$ constant)** | $Q_1$ = 0.6 L/hr | $Q_2$ = 3.5 L/hr | $Q_3$ = 9.6 L/hr |

**Table 3.** Simulation setup for three different sizes of cylindrical slice

This is indicated by the redness across the particle bed. However, a small number of particles might have less than the complete saturation. The RSD is measured from the water content of all the particles in all three cylindrical slices. Firstly the RSD is plotted as a function of time, as shown in Figure 12(A). It is observed that RSD reaches 0.1 at the end of spray for all sizes.

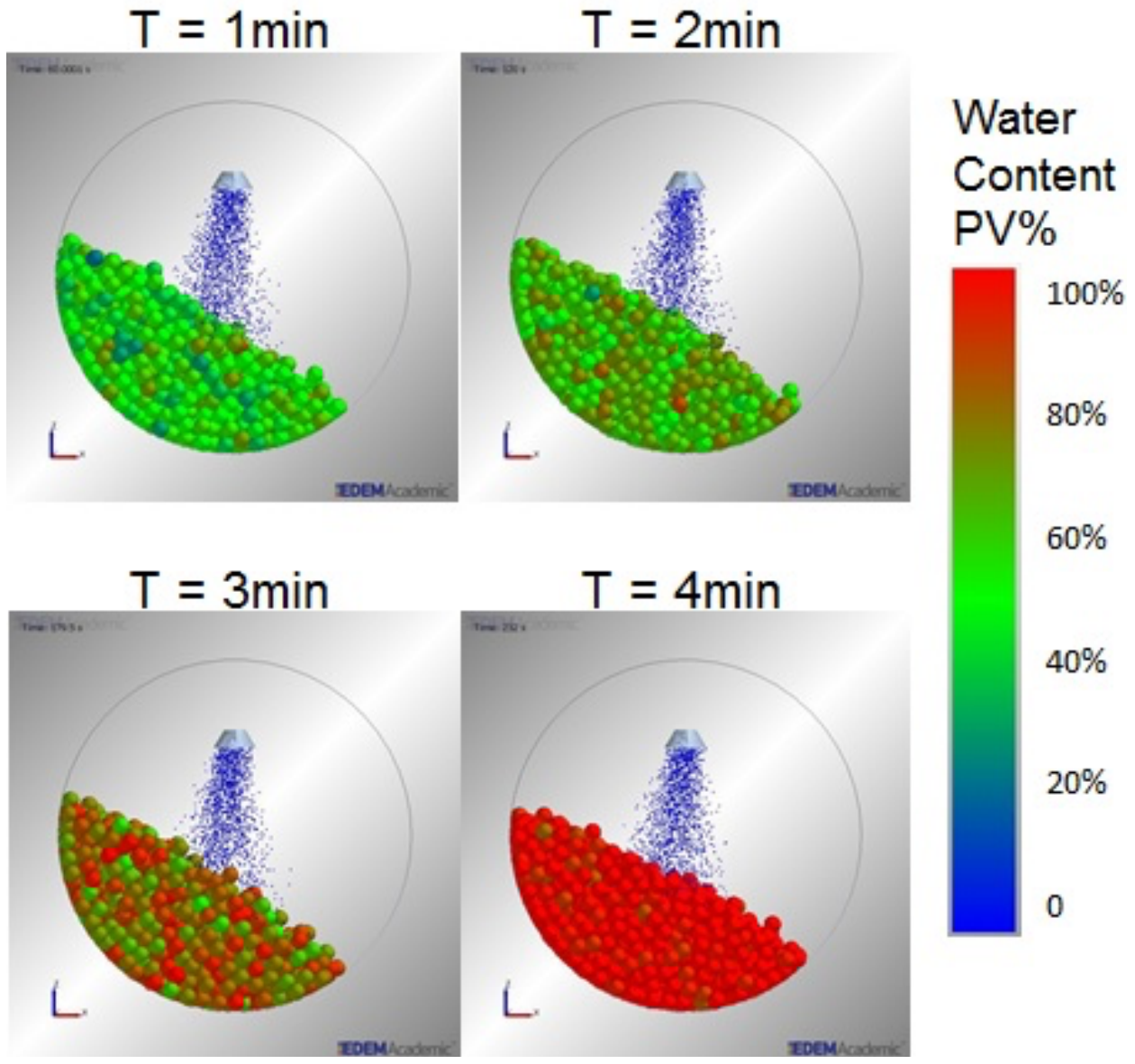

**Figure 11.** Simulation snapshots of a cylindrical slice (D=10cm) at different times.

The total amount of water needed to achieve complete pore filling is different for different sizes; as a result, the time to finish spray is also different. The spraying time is dependent on the amount of material and the spray rate. Then RSD is plotted as a function of the number of revolutions, as shown in Figure 12(B). Using the same Froude number and $C_Q$ number, the RSD curves collapse into one curve. Similar performance is observed in all three cylindrical slices. Thus the dimensionless numbers can be used for system scale-up. The defined four dimensionless numbers are verified to fully scale up the system.

## 5.2 Determination of the Operation Parameters

In this section, we explored how the impregnation process can be optimized, i.e., how to determine the operation parameters to achieve uniformity in the shortest time possible. The correlation between vessel size and the rotational speed of the vessel is given by the Froude number. For a given vessel size, $Fr$ is proportional to $\Omega^2$.

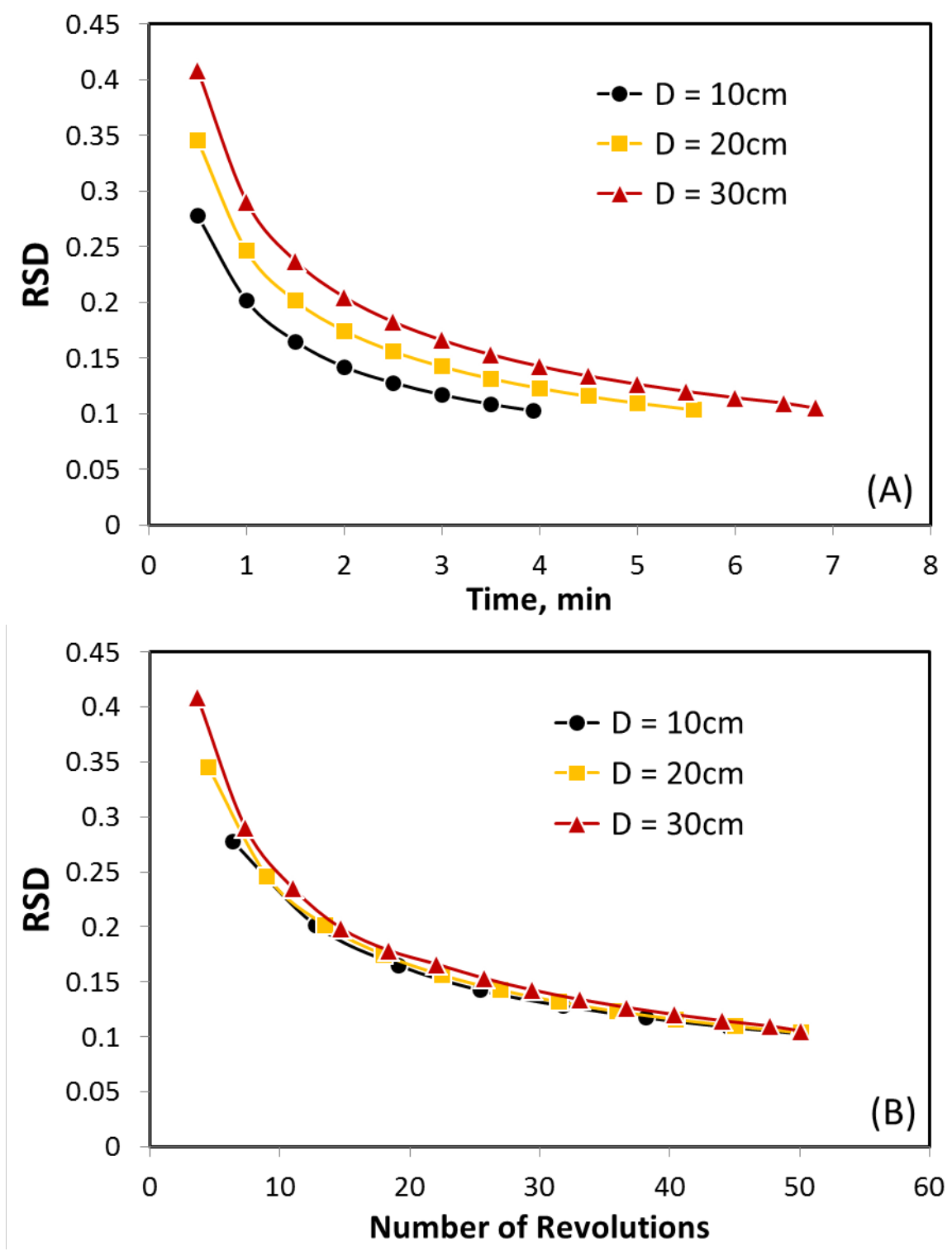

**Figure 12.** RSD plots for different cylinder sizes: (A) as a function of time and (B) as a function of the number of revolutions.

Within different $Fr$ regimes, different types of bed motion have been identified [46], such as slipping, rolling and cascading. The rolling bed is preferred, providing favorable conditions for mixing and uniformity. For a given vessel size, if the rotational speed and Froude number are too small, the system does not move, or the particle bed moves as if it rocked like a cradle, and no mixing occurs. At substantial rotational speeds and Froude numbers, the vessel rotates at high speed, causing particle collision and attrition, which are to be avoided during the impregnation process. An optimal regime of the Froude number can be found when the system shows the best mixing behavior with a flatbed profile and steady state.

In section 4.2, we performed a validation of the simulation parameters using our experiments performed in a full-length cylinder, which is computationally very costly with all the geometrical parameters and velocities matched 1:1. This allows us to get the constants $\kappa$ for water transfer and verify the other simulation parameters proposed in Table 1. Once the code is verified, it is more convenient to predict properties in a portion of the cylinder, a "cylindrical slice" of smaller length but the same diameter that takes much less computational time. DEM simulations were performed in a cylindrical slice with dimensions: D=20cm and W=8cm. The rotational speeds (Ω) were varied from (1rpm, 5rpm, 10rpm, 15rpm, 20rpm, 25rpm, and 50rpm) and the Froude number was calculated as $Fr = \Omega^2 D/2g$ . A constant spray rate of 3 L/hr was used. Figure 13 shows the simulation snapshots at the end of the spray for different rotational speeds. For the smallest rotational speed (1 rpm), a slipping motion was observed. In this regime, we can observe that the particle bed moved with the rotating wall up to a certain angle and then slid back as a whole on the wall surface. Minimum particle mixing took place in the slipping motion. Some of the particles were still dry or will little amount of water. This state of motion is not desirable in practice. For the intermediate rotational speed (10rpm-25rpm), particles on the surface of the bed flow in a uniform and static motion. The larger part of the bed was continuously transported upwards with the rotating wall.

The bed surface was observed to be nearly leveled, and the dynamic angle of repose depended both on the rotational speed and the fill level. This type of motion made a uniform and good intermixing possible. The particles showed uniform water content. For the largest rotational speed (50 rpm), a cascading bed motion was observed. The bed surface appeared arched, and the area of the wetting zone was reduced. It was also observed that some particles detached from the bed and were thrown off into the free space. Particles with large velocities tended to crash with each other, leading to particle collision and attrition, which are to be avoided during the impregnation process.

The relative standard deviation is then calculated from the water content in the axial direction of all the particles at the end of the spray. Figure 14 shows the RSD at completion (at the end of spray) vs. the logarithm of the Fr number. We can then obtain the value of the Fr number for which the RSD is minimal (Optimal Fr). The smallest RSD implies the best mixing and content uniformity. The data points are fitted by a 2nd-order polynomial, and when RSD is minimal, the optimal Froude number obtained is Fr = 0.009. It is worth noticing that the RSD corresponding to the points closer to the minimum is very close to each other. For that reason, we believe that there is a range of acceptable Froude numbers that would give an acceptable level of mixing and fluid content uniformity with values of RSD close to 0.1.

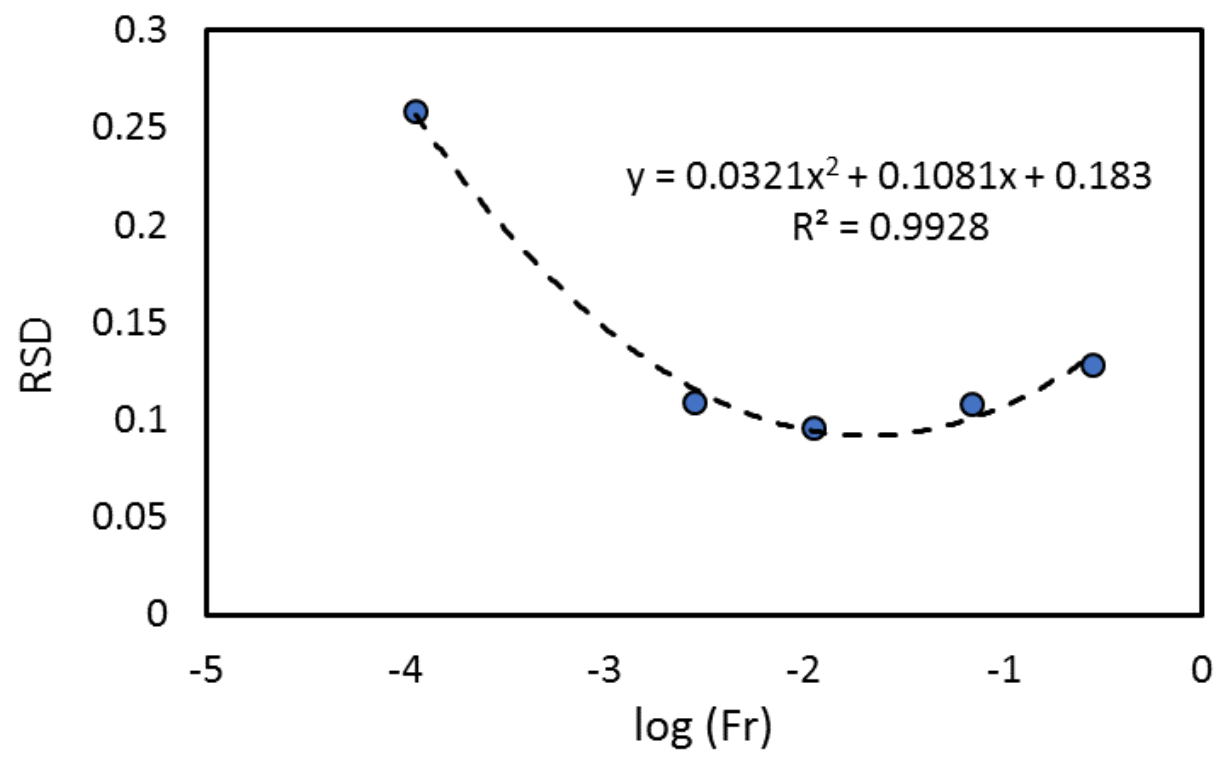

**Figure 14.** RSD as a function of the logarithm of Froude number. The value of optimal Fr is indicated by the minimal RSD.

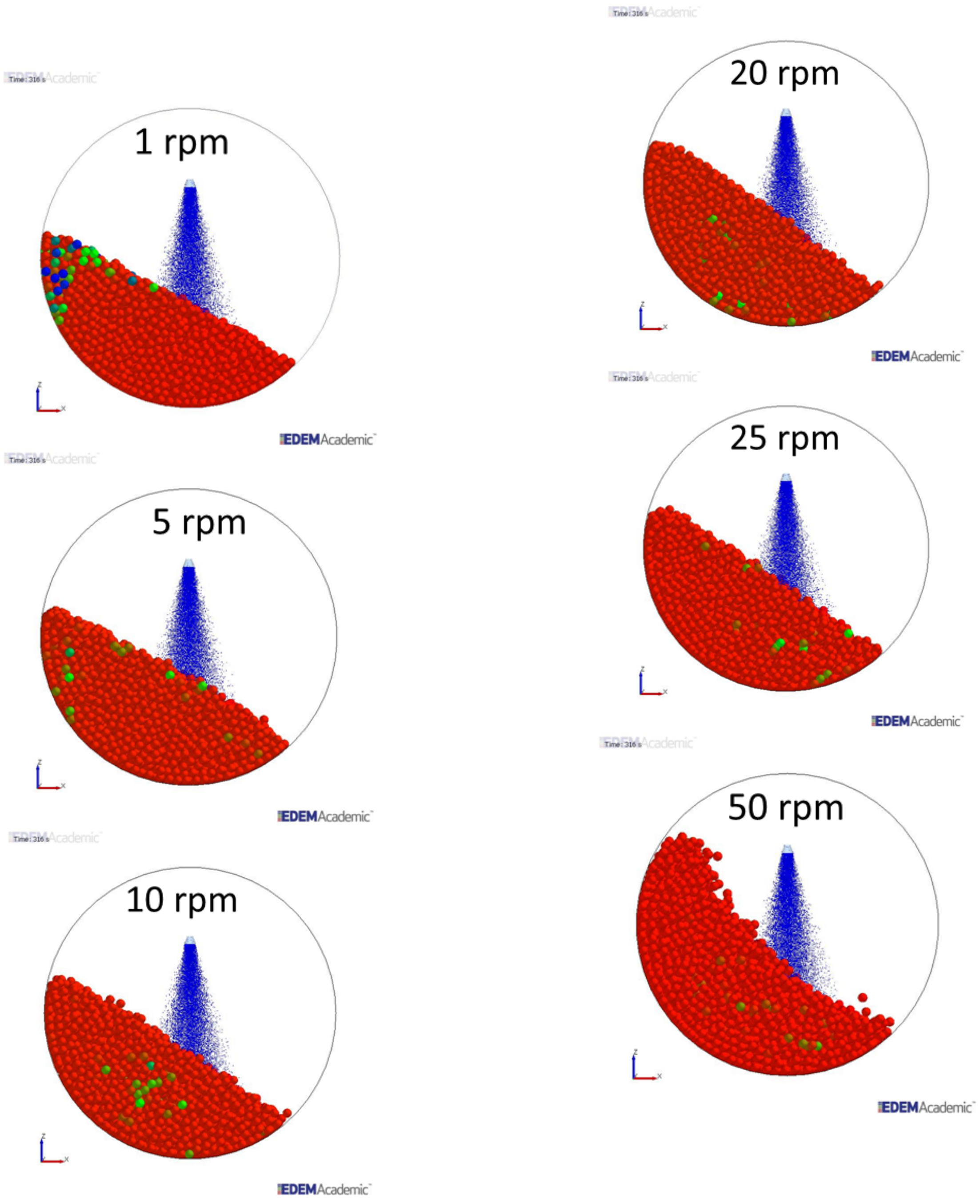

**Figure 13.** Simulation snapshots of a cylindrical slice (D=20cm) for different rotational speeds. Increasing the rotational speed from 1rpm to 25rpm, the particle bed maintains the rolling regime. However at higher speeds (50rpm), a cascading regime is observed.

Next, we examined the effect of particle size on the scale up performance. DEM simulations were performed in a cylindrical slice of D=10cm for different particle sizes (d= 2.5mm, 5mm, and 10mm).

The relative standard deviation is measured from the water content of all the particles in all three cylindrical slices. Figure 15(A) shows RSD plotted as a function of the number of revolutions for different particle sizes. RSD starts with a larger value for smaller particle sizes, due to the fact that mixing is poorer for smaller particles. Uniformity can be achieved eventually after the same number of revolutions.

(A)

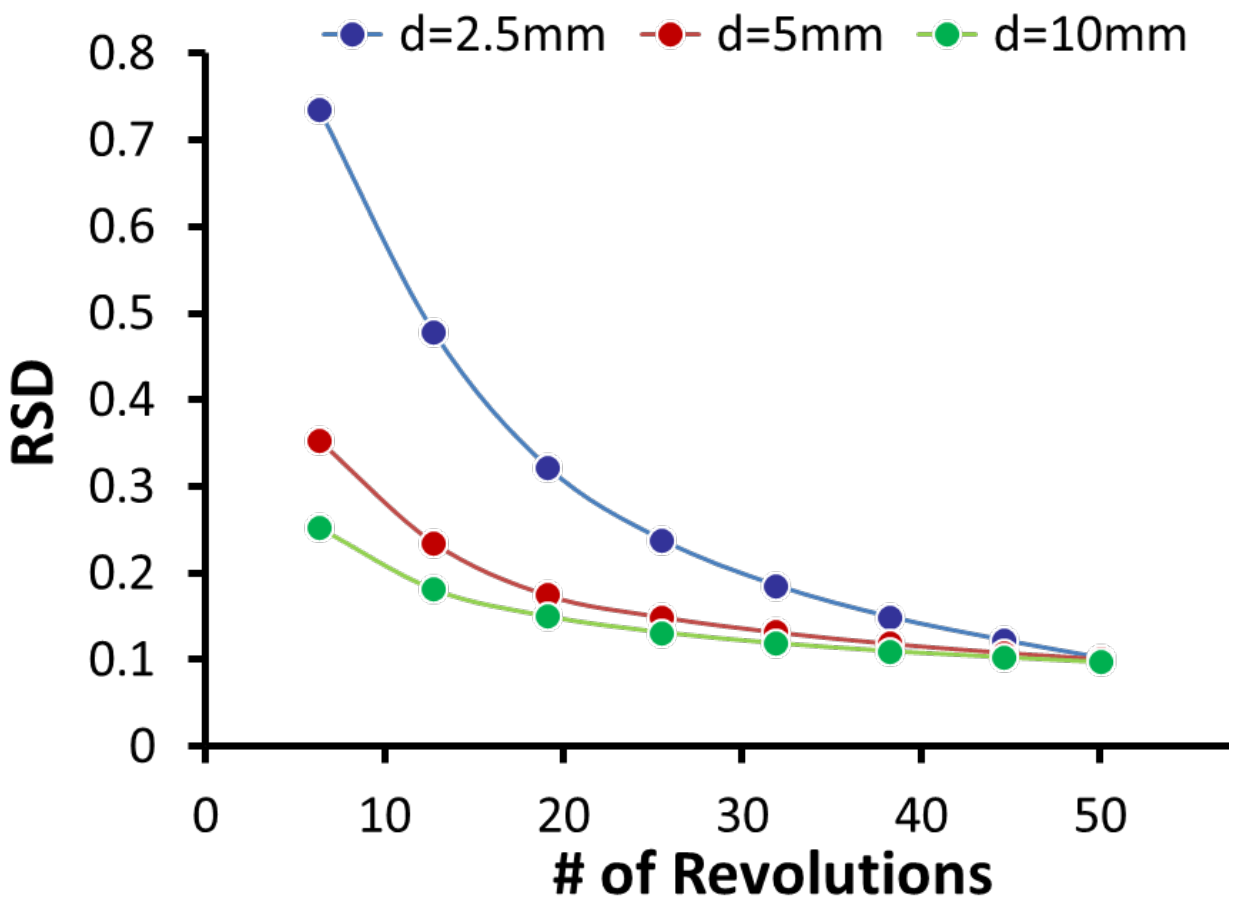

(B)

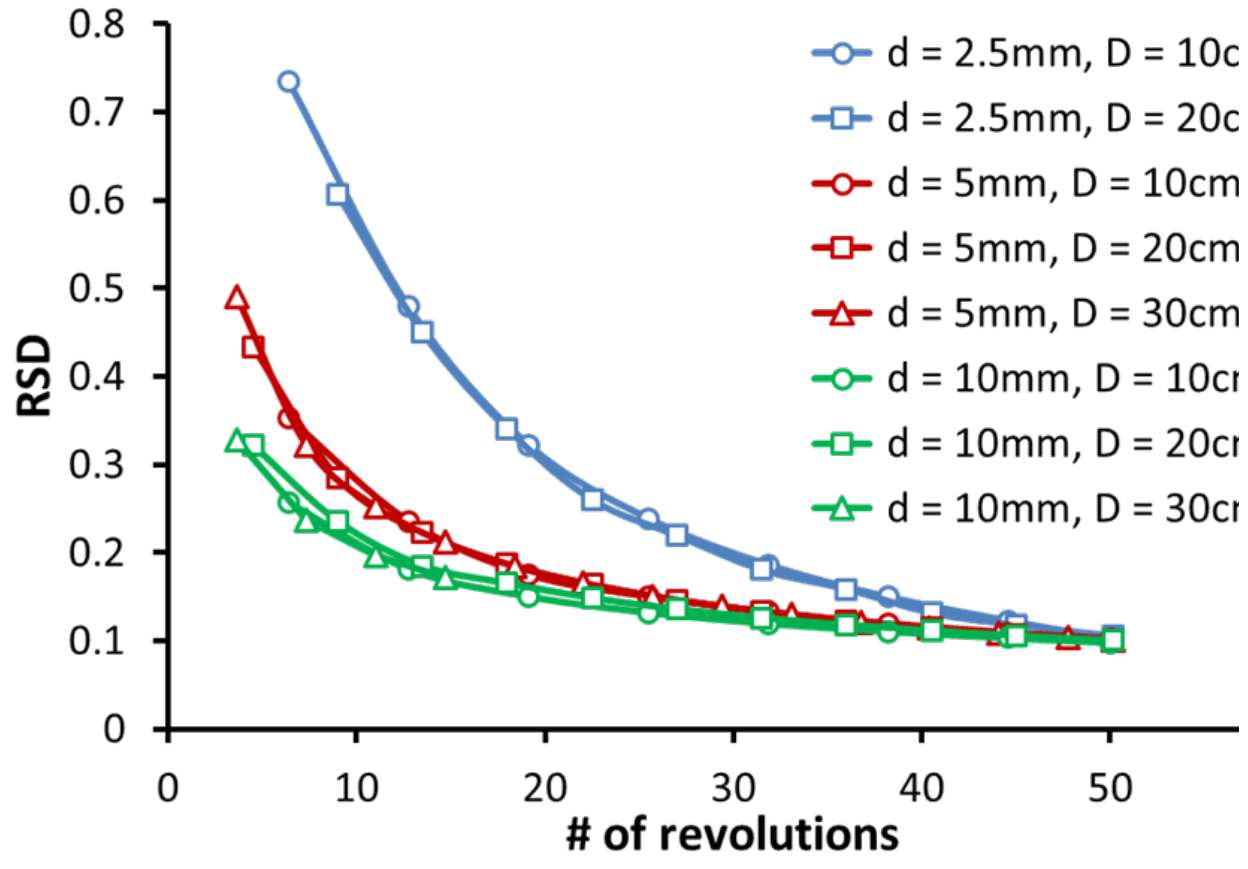

**Figure 15.** RSD plot as function of number of revolutions for different particle sizes for (A) fixed cylinder diameter (D = 10cm) and (B) different cylinder diameters.

To investigate the effect of particle size on scale up, DEM simulations were performed for other cylinder sizes (D= 20cm and 30cm) for the same range of particle sizes. In each particle size group, the operation conditions are determined by the scale-up rules. Figure 15(B) shows the RSD as a function of the number of revolutions for different D/d ratios (diameter of the vessel D to the diameter of the particle d); D/d = 10,20, 30,40,60,80,100, 120, 300.

We observe that for each particle size group, the relative standard deviation curves overlap with each other indicating that particle size does not have an effect. This suggests that particle size apparently does not have a significant effect on the scale up for the values of D/d considered. This, however, needs more clarification. We see that for the same vessel size, the RSD curves do not collapse hence, increasing the particle size does have an effect on RSD. The smaller the particle the larger the RSD, i.e. the poorer the fluid content uniformity. This is consistent with our previous publication for impregnation of particles of different sizes and morphologies [38]. To understand this better we plotted the RSD as a function of vessel diameter and see that the RSD saturates for ratios larger than D/d=10 and becomes independent of the vessel diameter D. This suggests that when the cylinder diameter is large enough (at least 10 times larger than the particle diameter d) the curves overlap and the particle size does not have an effect in scale up. However this also suggest that when the ratio D/d less than 10 the particle size may have an effect on scale up and needs to be taken into account.

The critical value for D/d has not been determined in this study, hence more works needs to be done. Notice that we have not used the size of the particles in any of the proposed dimensionless numbers. There should be a ratio D/d for which particle size d has an effect in scale-up and needs to be considered in a dimensionless number. Figure 16 shows snapshots of the initial setup of these studies.

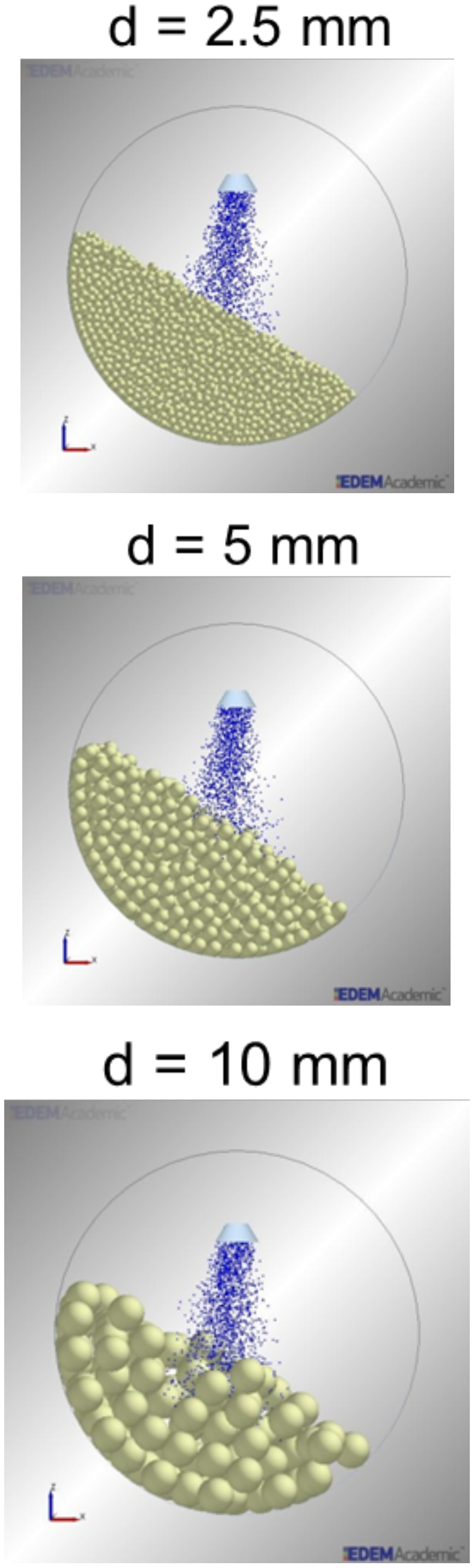

**Figure 16.** Initial simulation setup of different particle sizes. The cylinder diameter is D = 10cm.

## 7. Conclusions

We developed a systematic and general approach to scale up the impregnation process. The defined five dimensionless numbers fully scale up the system. We performed impregnation experiments in both a small and a large cylinder. Our selected Froude number lies within the range of the optimal Froude predicted by the simulations. The five selected dimensionless numbers give excellent scale-up performance since curves overlap, indicating similarity (geometric, kinematic, and dynamic) and hence similar performances for all different sizes. The scale-up method has been validated using both simulations and experiments. The scale-up rules were also applied to different particle sizes in simulations and showed good reproducibility.

Our results show that when the diameter of the vessel is large enough concerning the diameter of the particle, the particle size is not relevant in the scale-up of the system. There is a critical size for which particle size does have an effect on scale up and that is when D/d is within a range of values less than 10. The critical value has not been determined in this study.
Notice also that our results were obtained from a limited number of process conditions, two sizes of cylindrical vessels, and only one type of material with a unimodal particle size distribution. The final product quality is mainly determined by mixing and dispersion. Issues may be caused when applying the proposed dimensionless quantities to a cohesive material that mixes poorly due to the formation of agglomerates.

It would also be interesting to examine different particle size distributions in future work.

## Acknowledgments

The authors would like to acknowledge the Catalyst Manufacturing Consortium at Rutgers University for funding and support. We thank Deval Sharma and Matthew Borsellino for their assistance with the experiments. NSF DMR-1945909 provided partial funding.

## Author contributions

YS performed the simulations and experiments. BB advised on the project. BA and HAM contributed to the analysis of simulations, and MST performed DEM simulations, wrote the paper, and directed the project.

## References

1. Schwarz, J.A., C. Contescu, and A. Contescu, *METHODS FOR PREPARATION OF CATALYTIC MATERIALS.* Chemical Reviews, 1995. **95**(3): p. 477-510.
2. van Dillen, A.J., et al., *Synthesis of supported catalysts by impregnation and drying using aqueous chelated metal complexes.* Journal of Catalysis, 2003. **216**(1-2): p. 257-264.
3. Barthe, L., et al., *Synthesis of supported catalysts by dry impregnation in fluidized bed.* Chemical Engineering Research & Design, 2007. **85**(A6): p. 767-777.
4. Munnik, P., P.E. de Jongh, and K.P. de Jong, *Recent Developments in the Synthesis of Supported Catalysts.* Chemical Reviews, 2015. **115**(14): p. 6687-6718.
5. Rossetti, I. and M. Compagnoni, *Chemical reaction engineering, process design and scale-up issues at the frontier of synthesis: Flow chemistry.* Chemical Engineering Journal, 2016. **296**: p. 56-70.
6. Johnstone, R.E. and M.W. Thring, *Pilot Plants, Models and Scale-up Methods in Chemical Engineering*. 1957.
7. Elson, T., *Scale-up in Chemical Engineering.* , in *Concepts of Chemical Engineering*, S.J.R. Simons, Editor. 2007, Royal Society of Chemistry.
8. Zlokarnik, M., *Dimensional Analysis and Scale-up in Chemical Engineering*. 1991: Springer-Verlag.
9. Zlokarnik, M., *Problems in the application of dimensional analysis and scale-up of mixing operations.* Chemical Engineering Science, 1998. **53**(17): p. 3023-3030.
10. Couper, J.R., J.R. Fair, and W.R. Penney, *Mixing and agitation.* Chemical Process Equipment: Selection and Design, 2010: p. 273-324.
11. Levin, M., *How to Scale Up Scientifically.* Pharmaceuetical Technology, 2005. **2005 Supplement**(1): p. S4-S11.
12. Wang, R.H. and L.T. Fan, *Methods for scaling-up tumbling mixers.* Chemical Engineering, 1974. **81**(11): p. 88-94.
13. Muzzio, F.J. and A.W. Alexander, *Scale Up of Powder Blending Operations.* Pharmaceuetical Technology, 2005. **2005 Supplement**(1): p. S34-S43.
14. Litster, J.D., et al., *Liquid distribution in wet granulation: dimensionless spray flux.* Powder Technology, 2001. **114**(1-3): p. 32-39.
15. Hapgood, K.P., et al., *Dimensionless spray flux in wet granulation: Monte-Carlo simulations and experimental validation.* Powder Technology, 2004. **141**(1-2): p. 20-30.
16. Boerefijn, R., P.Y. Juvin, and P. Garzon, *A narrow size distribution on a high shear mixer by applying a flux number*

*approach.* Powder Technology, 2009. **189**(2): p. 172-176.
17. Pandey, P. and S. Badawy, *A quality by design approach to scale-up of high-shear wet granulation process.* Drug Development and Industrial Pharmacy, 2016. **42**(2): p. 175-189.
18. Turton, R. and X.X. Cheng, *The scale-up of spray coating processes for granular solids and tablets.* Powder Technology, 2005. **150**(2): p. 78-85.
19. Pandey, P., M. Katakdaunde, and R. Turton, *Modeling weight variability in a pan coating process using Monte Carlo simulations.* Aaps Pharmscitech, 2006. **7**(4).
20. Kariuki, W.I.J., et al., *Distribution nucleation: Quantifying liquid distribution on the particle surface using the dimensionless particle coating number.* Chemical Engineering Science, 2013. **92**: p. 134-145.
21. Poozesh, S. and E. Bilgili, *Scale-up of pharmaceutical spray drying using scale-up rules: A review.* International Journal of Pharmaceutics, 2019. **562**: p. 271-292.
22. Litster, J.D., *Scaleup of wet granulation processes: science not art.* Powder Technology, 2003. **130**(1-3): p. 35-40.
23. Pandey, P., et al., *Scale-up of a pan-coating process.* Aaps Pharmscitech, 2006. **7**(4).
24. Joglekar, A., et al., *Mathematical model to predict coat weight variability in a pan coating process.* Pharmaceutical Development and Technology, 2007. **12**(3): p. 297-306.
25. Mueller, R. and P. Kleinebudde, *Comparison of a laboratory and a production coating spray gun with respect to scale-up.* Aaps Pharmscitech, 2007. **8**(1).
26. Prpich, A., et al., *Drug product modeling predictions for scale-up of tablet film coating-A quality by design approach.* Computers & Chemical Engineering, 2010. **34**(7): p. 1092-1097.
27. Just, S., et al., *Optimization of the inter-tablet coating uniformity for an active coating process at lab and pilot scale.* International Journal of Pharmaceutics, 2013. **457**(1): p. 1-8.
28. Toschkoff, G. and J.G. Khinast, *Mathematical modeling of the coating process.* International Journal of Pharmaceutics, 2013. **457**(2): p. 407-422.
29. Agrawal, A.M. and P. Pandey, *Scale Up of Pan Coating Process Using Quality by Design Principles.* Journal of Pharmaceutical Sciences, 2015. **104**(11): p. 3589-3611.
30. Boehling, P., et al., *Simulation of a tablet coating process at different scales using DEM.* European Journal of Pharmaceutical Sciences, 2016. **93**: p. 74-83.
31. Boehling, P., et al., *Analysis of large-scale tablet coating: Modeling, simulation and experiments.* European Journal of Pharmaceutical Sciences, 2016. **90**: p. 14-24.
32. Dennison, T.J., et al., *Design of Experiments to Study the Impact of Process Parameters on Droplet Size and Development of Non-Invasive Imaging Techniques in Tablet Coating.* Plos One, 2016. **11**(8).
33. Suzuki, Y., et al., *A Novel Scale Up Model for Prediction of Pharmaceutical Film Coating Process Parameters.* Chemical & Pharmaceutical Bulletin, 2016. **64**(3): p. 215-221.
34. Ban, J., et al., *Scaling Inter-Tablet Coating Variability in a Horizontal Rotating Drum.* Aiche Journal, 2017. **63**(9): p. 3743-3755.
35. Chaudhuri, B., et al., *Cohesive effects in powder mixing in a tumbling blender.* Powder Technology, 2006. **165**(2): p. 105-114.
36. Nakamura, H., H. Fujii, and S. Watano, *Scale-up of high shear mixer-granulator based on discrete element analysis.*

Powder Technology, 2013. **236**: p. 149-156.
37. Romanski, F.S., et al., *Dry catalyst impregnation in a double cone blender: A computational and experimental analysis.* Powder Technology, 2012. **221**: p. 57-69.
38. Shen, Y.Y., W.G. Borghard, and M.S. Tomassone, *Discrete element method simulations and experiments of dry catalyst impregnation for spherical and cylindrical particles in a double cone blender.* Powder Technology, 2017. **318**: p. 23-32.
39. Ding, Y.L., et al., *Scaling relationships for rotating drums.* Chemical Engineering Science, 2001. **56**(12): p. 3737-3750.
40. Cundall, P.A. and O.D.L. Strack, *A discrete numerical model for granular assemblies.* Geotechnique, 1979. **29**.
41. Hertz, H., *On the contact of elastic solids.* Journal fur die reine und angewandte Mathematik 1882. **92**: p. 156-171.
42. Tsuji, Y., T. Tanaka, and T. Ishida, *LAGRANGIAN NUMERICAL-SIMULATION OF PLUG FLOW OF COHESIONLESS PARTICLES IN A HORIZONTAL PIPE.* Powder Technology, 1992. **71**(3): p. 239-250.
43. Mindlin, R.D. and H. Deresiewicz, *Elastic spheres in contact under varying oblique forces.* Journal of Applied Mechanics, 1953. **20**.
44. *Guidance for Industry: Powder Blends and Finished Dosage Units — Stratified In-Process Dosage Unit Sampling and Assessment*. 2003; Available from: http://academy.gmp-compliance.org/guidemgr/files/5831DFT.PDF.
45. *<905> Uniformity of Dosage Units, The United States Pharmacopeial Convention*. 2011; Available from: https://www.usp.org/sites/default/files/usp/document/harmonization/gen-method/q0304_stage_6_monograph_25_feb_2011.pdf.
46. Mellmann, J., *The transverse motion of solids in rotating cylinders - forms of motion and transition behavior.* Powder Technology, 2001. **118**(3): p. 251-270.